\pdfoutput=1

\documentclass[prl,twocolumn,amsmath,amssymb,floatfix,superscriptaddress]{revtex4}
\usepackage{graphicx}
\usepackage{dcolumn}
\usepackage{bm}
\usepackage{amsmath}
\usepackage{amssymb,xcolor,soul}

\usepackage[normalem]{ulem}
\usepackage[mathscr]{euscript}
\begin{document}

\title{First Principles Calculation of Dzyaloshinskii-Moriya Interaction: A Green's function Approach }

\author{Farzad Mahfouzi}
\email{Farzad.Mahfouzi@gmail.com}
\affiliation{Department of Physics and Astronomy, California State University, Northridge, California 91330-8268, USA}
\author{Nicholas Kioussis}
\email{nick.kioussis@csun.edu }
\affiliation{Department of Physics and Astronomy, California State University, Northridge, California 91330-8268, USA}

\begin{abstract}
We present a Greens function approach to calculate the  Dzyaloshinskii-Moriya interactions (DMI) from first principles electronic structure calculations, that is computationally more efficient and accurate than  the most-commonly employed supercell and generalized Bloch-based approaches. The method is applied to the (111) Co/Pt bilayer where the Co- and/or Pt-thickness dependence of the DMI coefficients are calculated. Overall, the calculated DMI are in relatively good agreement with the corresponding values reported experimentally. Furthermore, we investigate the effect of strain in the DMI tensor elements and show that the isotropic N\'{e}el DMI can be significantly modulated by the normal strains, $\epsilon_{xx},\epsilon_{yy}$ and is relatively insensitive to the shear strain, $\epsilon_{xy}$. Moreover, we show that anisotropic strains, $(\epsilon_{xx}-\epsilon_{yy})$ and 
 $\epsilon_{xy}$, result  in the emergence of anisotropic  N\'{e}el- and Bloch-type DMIs, respectively.
			
\end{abstract}
 \date{\today}

\maketitle

\section{Introduction}\label{sec:intro}
Magnetic Skyrmions\cite{Fert2013,Kiselev2011,Nagaosa2013,Bogdanov1989,Bogdanov1994} are noncollinear spin configurations found in materials with strong spin-orbit coupling (SOC) and non-centrosymmetric crystal structures which lack an inversion center. 
Their existence were theoretically predicted three decades ago\cite{Bogdanov1989,Bogdanov1994} and experimentally observed 
recently in chiral magnets and other B20-type bulk materials, such as 
MnSi\cite{Muhlbauer2009}, Fe$_{1-x}$Co$_x$Si\cite{Yu2010}, FeGe\cite{Yu2011} and Mn$_{1-x}$Fe$_x$Ge\cite{Shibata2013}. 
Bilayer devices consisting of ferromagnets interfaced with heavy-metals (HMs), with their mirror symmetry 
broken across the interface, also provide the necessary condition for the emergence of skyrmions.\cite{Fert2017,Everschor2018,Finocchio2016,Soumyanarayanan2017,Jiang2015,Hrabec2017,Sampaio2013,Upadhyaya2016,Bttner2017,HellmanRMP2017} The noncollinearity of the spin configuration is conventionally parameterized the by Dzyaloshinskii-Moriya interaction\cite{Del1958,Moriya1960} (DMI); an anti-symmetric exchange interaction between atomic magnetic moments, which is generally measured experimentally using Brillouin Light Scattering (BLS) \cite{Udvardi2009,Costa2010,Moon2013,Volkov2019,Di2015,Ma2016,Ma2017}. 
The DMI energy for two spins $\vec{S_1}$ and $\vec{S_2}$ separated by a distance $\vec{R}$  is of the form $E_{DMI} = \vec{d}_{\vec{R}}\cdot(\vec{S}_1\times\vec{S}_2)$,\cite{Del1958,Moriya1960} where the DMI vector generally consists of two terms\cite{Kim2018}, referred to as the Bloch, $\vec{d}^B$ and N\'{e}el, $\vec{d}^N$ components, respectively. In metallic systems the DMI is mediated by the conduction electrons. 
Conduction electrons with spin-momentum coupling of the Rashba type are known to lead to a N\'{e}el DMI. \cite{Yang2018,Manchon2015,Kezsmarki2015,Kurumaji2017} This type of helicity is often present at the interface between two different materials, where, the broken inversion symmetry normal to the interface results in a DMI vector of the form, $\vec{d}^{N}=d^{N}\vec{e}_z\times\vec{R}/|\vec{R}|$, 
where $\vec{e}_z$ is the unit vector normal to the interface.
On the other hand, systems with Dresselhouse type SOC, which is often manifested in the bulk region, exhibit a Bloch type DMI vector,\cite{Muhlbauer2009,Seki2010,Onose2010} of the form $\vec{d}^{B}=-d^{B}\vec{R}/|\vec{R}|$.

Magnetic skyrmions are promising candidates for several applications,\cite{Finocchio2019} including racetrack memory\cite{Tomasello2014,Zhang2015_1,Kang2016}, skyrmion based logic gates,\cite{Zhang2015_2} microwave devices\cite{Finocchio2015}, and neuromorphic computing\cite{Li2018,He2017}. Such devices require efficient ways of skyrmion manipulation \cite{Romming2015,Koshibae2014,Rowland2015,Daz2016,Masell2020,Buttner2017}, including creation, annihilation, motion, and detection. 
Interestingly, recent experiments have demonstrated the electric field-\cite{Koyama2018,Zhang2018,Srivastava2018} 
and/or strain-mediated\cite{Gusev2020} modulation of the interfacial DMI which may provide a novel function for skyrmion- or chiral domain wall-based spintronic devices. However, in order for the skyrmion nucleation to be energetically feasible and stable at room temperature, large DMI values are required,\cite{Boulle2016,Woo2016,Soumyanarayanan2017,Moreau2016} making it difficult to significantly modulate the sign and/or amplitude of the DMI. To address this issue, first principles calculations provide a promising avenue in the search for devices with large DMI that are also responsive to external perturbations.

Several first principles approaches have so far been introduced to calculate the DMI. These include the super-cell \cite{ChshievDMI2015}, the generalized Bloch theorem \cite{Zimermann2019}, the Korringa-Kohn-Rostoker
 Green function (KKR-GF) \cite{Mankowsky2017}  and the Berry curvature \cite{Frank2017} methods. Even though the super-cell approach is the most popular to evaluate the DMI from first principles,  it is the most computationally demanding and hence limits its application 
 to systems with small number of atoms and only to DMIs between nearest-neighbor atoms. 
 The generalized Bloch theorem approach is computationally less demanding than the super-cell, albeit, at the cost of taking into account the effect of SOC perturbatively. 
 On the other hand, the KKR-GF and Berry curvature methods are computationally the most efficient while taking into account the effect of SOC exactly. 
 
Developing an accurate and efficient first principles approach to calculate the DMI and its manipulation via current\cite{Kato2019}, 
electric field\cite{Yang2018_1}, and strain\cite{Deger2020,Udalov2020} is of paramount importance  in the spintronics community. 
 In this paper, we employ Greens function method to calculate the atomistic exchange coupling from the  magnon band structure in the spin-spiral regime. The DMIs are calculated for the Pt/Co bilayer for various Pt and/or Co thickness and are compared with the corresponding values reported in experiments, where the overall agreement is good. Furthermore, we have investigated the effect of strain on the DMI tensor elements which is in relative agreement with the recent experimental observation.\cite{Gusev2020}

\section{Theoretical formalism}\label{sec:GenForm}
\subsection{Phenomenological Modeling}
Investigations of magnetic properties of materials at the
phenomenological level based on model spin Hamiltonians
may be formulated either in the continuum (micromagnetic) or in the discrete atomistic representation.

The Hamiltonian governing the atomic spin dynamics can be generally written as,
 \begin{align}
 E_{M}&=-\frac{1}{2}\sum_{ij;\vec{R}\vec{R}';\alpha\beta}J^{ij,\alpha\beta}_{\vec{R}-\vec{R}'}\vec{m}^i_{\vec{R}}\cdot\vec{e}_{\alpha}\vec{m}^{j}_{\vec{R}'}\cdot\vec{e}_{\beta},
 \end{align}
where $\vec{m}^{i(j)}_{\vec{R}}$ is a unit vector along the local magnetic moment of the $i$ ($j$)th atom within the unit cell at $\vec{R}$ ($\vec{R}'$), $J^{ij,\alpha\beta}_{\vec{R}-\vec{R}'}$ is the exchange coupling between the two local moments, and $\vec{e}_{\alpha}$ is a unit vector along the 
$\alpha=x,y,z$ direction.

The Landau-Lifshitz equation of motion is then given by,
\begin{subequations}
	\begin{align}
	\frac{M_i}{2\mu_B}\frac{\partial}{\partial t}\vec{m}^i_{\vec{R}}&=\vec{m}^i_{\vec{R}}\times[\hat{J}\vec{m}]^i_{\vec{R}}, \\	[\hat{J}\vec{m}]^{i}_{\vec{R}}\cdot\vec{e}_{\alpha}&=\sum_{\vec{R}'j\beta}J^{ij,\alpha\beta}_{\vec{R}-\vec{R}'}\vec{m}^{j}_{\vec{R}'}\cdot\vec{e}_{\beta},
	\end{align}		
\end{subequations}
where $M_i$ is the magnetic moment of the $i$th atom. In the linear regime the time-dependent local moment direction can be written as 
\begin{align}
\vec{m}^i_{\vec{R}}(t)&\approx\vec{e}_z+\delta^{i,x}_{\vec{R}}(t)\vec{e}_x+\delta^{i,y}_{\vec{R}}(t)\vec{e}_y
\end{align}	
where $\vec{e}_z$ is the easy axis of the magnetization.  

 In this case, the resulting equations of motion in the spin-spiral regime are given by,
\begin{subequations}\label{eq:linearEoM}
\begin{align}
\frac{M_{tot}}{2\mu_B}\frac{\partial \delta^x_{\vec{q}}}{\partial t}&=-{J}^{yx}_{\vec{q}}\delta^x_{\vec{q}}-({J}^{yy}_{\vec{q}}-{J}^{zz}_{0})\delta^y_{\vec{q}}\\
\frac{M_{tot}}{2\mu_B}\frac{\partial \delta^y_{\vec{q}}}{\partial t}&={J}^{xy}_{\vec{q}}\delta^y_{\vec{q}}+({J}^{xx}_{\vec{q}}-{J}^{zz}_{0})\delta^x_{\vec{q}},
\end{align}	
\end{subequations}
where, $M_{tot}=\sum_iM_i$ is the total magnetic moment per unit cell and 
\begin{align}
{J}_{\vec{q}}^{\alpha\beta}&=\sum_{ij,\vec{R}}{J}_{\vec{R}}^{ij,\alpha\beta}e^{i(\vec{r}_i-\vec{r}_{j}+\vec{R})\cdot\vec{q}}.
\end{align}

We can in turn make a connection between the parameters of the continuum model and the micromagnetic energy defined as 
\begin{align}
E_{M}=\sum_{\alpha\beta}\int d^3\vec{R}&\left[
A_{\alpha\beta}\frac{\partial {m}^{\alpha}}{\partial R_{\beta}}\frac{\partial {m}^{\alpha}}{\partial R_{\beta}}+K_{\alpha\beta}m^{\alpha}m^{\beta}\right. \nonumber\\
&\left.+D_{\alpha\beta}\Big(\vec{m}\times\frac{\partial \vec{m}}{\partial R_{\alpha}}\Big)\cdot\vec{e}_{\beta}\right] ,
\end{align}	
where  $A_{\alpha\beta}$, $K_{\alpha\beta}$ and $D_{\alpha\beta}$ are the exchange stiffness, MCA and DMI tensor elements, respectively.  
In this case the dynamical equations of motion are given by
\begin{align}\label{eq:ContModel}
\frac{M_{tot}}{4\mu_BV_{M}}\frac{\partial\vec{m}}{\partial t}=\sum_{\alpha\beta}\vec{m}\times&\left(-A_{\alpha\beta}\frac{\partial^2\vec{m}}{\partial R_{\alpha}\partial R_{\beta}}+\right.\\
&\left.K_{\alpha\beta}\vec{e}_{\alpha}m^{\beta}+D_{\alpha\beta}\frac{\partial\vec{m}}{\partial R_{\beta}}\times\vec{e}_{\alpha}\right).\nonumber
\end{align}	

Comparison of the linearized Eqs.~\eqref{eq:ContModel} with the system of Eqs.~\eqref{eq:linearEoM} yields 

\begin{subequations}\label{eq:ContEq}
\begin{align}
K_{\alpha\beta}&=\left.\frac{1}{2V_M}
 J^{\alpha\beta}_{\vec{q}}\right|_{\vec{q}=\Gamma},\label{eq:ContEqa}\\
D_{\alpha\beta}&=\left.\frac{1}{4V_M}Im\left(\sum_{\gamma\gamma'}\epsilon_{\beta\gamma\gamma'}\frac{\partial J^{\gamma\gamma'}_{\vec{q}}}{\partial q_{\alpha}}\right)\right|_{\vec{q}=\Gamma},\label{eq:ContEqb}\\
A_{\alpha\beta}&=\left.\frac{1}{2V_M}\frac{\partial^2 J^{\alpha\alpha}_{\vec{q}}}{\partial q_{\beta}^2}\right|_{\vec{q}=\Gamma}\label{eq:ContEqc}.
\end{align}	
\end{subequations}

Here, the Levi-Civita symbol, $\epsilon_{\alpha\beta\gamma}$ , is introduced to convert the third-order DMI tensor in Lifshitz invariant presentation \cite{Ado2020} to a second-order tensor and $V_M$ is the volume of the unit cell of the ferromagnetic film. Eq.~\eqref{eq:ContEqb} can be used to calculate the DMI tensor elements, $D_{\alpha\beta}$. However, consistent with the BLS method employed in experiments, 
in our first principles calculations we use the magnon dispersion method to evaluate the different DMI components. 

In the following we provide a phenomenological treatment of the DMI. This can be achieved by extending the N\'{e}el, $\vec{d}^{N}$ , and Bloch, $\vec{d}^{B}$ , components of the DMI vector for a pair of magnetic moments 
to a lattice of magnetic ions, the DMI tensor elements for a ferromagnetic/heavy metal bilayer with broken mirror symmetry along the $z$-axis, are given by
\begin{align}\label{eq:PhenDMI}
D_{\alpha\beta}&=\frac{1}{2V_M}\sum_{\vec{R}}\left[d^{N}_{\vec{R}}\frac{\vec{e}_{z}\times\vec{R}}{|\vec{R}|}+d^{B}_{\vec{R}}\frac{\vec{R}}{|\vec{R}|}\right]\cdot\vec{e}_{\alpha}R_{\beta}.
\end{align}	

For an isotropic system, where the interatomic DMI coupling depends only on $|\vec{R}|$, the  N\'{e}el and Bloch  components of the DMI tensor 
result in the off-diagonal and diagonal  matrix elements, respectively,   
\begin{align}\label{eq:Eq14}
D^{iso}_{\alpha\beta}&=D^N\vec{e}_{z}\cdot\vec{e}_{\alpha}\times\vec{e}_{\beta}+D^{B}\vec{e}_{\alpha}\cdot\vec{e}_{\beta},
\end{align}	
where, $D^N=\sum_{n}N_nR_nd^{N}_n/4V_M$ and $D^B=\sum_{n}N_nR_nd^{B}_n/4V_M$, and $N_n$ denotes 
the number of nearest-neighbors of the {\it n}th shell. 
For example for a film with in-plane hexagonal crystal structure we have,
$(R_n/a)^2=(1,3,4,7,9,...)$ and $N_n=(6,6,6,12,6,...)$. 

\begin{figure} [tbp]
	{\includegraphics[scale=0.32,angle=0,trim={0cm 4cm 0.0cm 2.0cm},clip,width=0.5\textwidth]{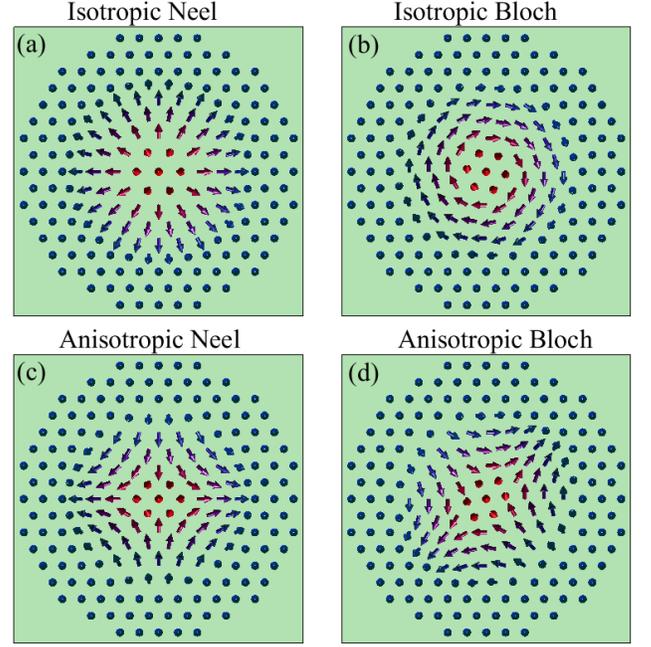}}%
	\caption{Schematic spin texture for skyrmions resulting from isotropic (a) N\'{e}el and (b) Bloch DMI tensor components, and anti-skyrmions originating from anisotropic (c)N\'{e}el  and (d) Bloch DMI tensor components. The red (blue) color of the arrows denotes the direction of the local magnetic moment along $+\vec{e}_z$ ($-\vec{e}_z$) direction. }
	\label{fig:fig1}
\end{figure}

On the other hand, for an anisotropic system, the diagonal elements, $D_{\alpha\alpha}$, of the DMI matrix are not necessarily identical and the off-diagonal elements (apart from the antisymmetric N\'{e}el component) can additionally acquire symmetric components. 
In this case, in line with Eq.~\eqref{eq:Eq14} the {\it average} value of the diagonal elements denotes the {\it isotropic} Bloch DMI, while the additional terms responsible for the difference in the diagonal elements denote the {\it anisotropic} Bloch DMI.
Similarly, we refer to the symmetric off-diagonal components of the DMI matrix as the {\it anisotropic} N\'{e}el components. 
For a bilayer system with broken mirror symmetry along the $z$-axis and two-fold ($C_{2v}$) in-plane rotational symmetry, the nonzero DMI tensor elements include, $D_{xx}$, $D_{xy}$, $D_{yx}$, and $D_{yy}$. In this case, the four DMI components are, $D^{N}_{iso}=(D_{xy}-D_{yx})/2$, $D^{B}_{iso}=(D_{xx}+D_{yy})/2$, $D^{N}_{an}=(D_{xy}+D_{yx})/2$, and $D^{B}_{an}=(D_{xx}-D_{yy})/2$, which in turn give rise to 
different types of skyrmions. 
In Figs.~\ref{fig:fig1}(a) and (b) we show the N\'{e}el and Bloch type skyrmions, respectively, corresponding to the 
{\it isotropic}  N\'{e}el and Bloch DMIs. On the other hand, as dispayed in Figs.~\ref{fig:fig1}(c) and (d), 
the {\it anisotropic}   and N\'{e}el Bloch DMIs,\cite{Camosi} which originate from  the anisotropy in the planar crystal structure 
 give rise to anti-skyrmions\cite{Zimmermann2017} that are related to each 
other by a 45$^{o}$ in-plane rotation. 

\subsection{Magnon Dispersion}
 Solving the linearized LLG equations of motion, Eqs.~\eqref{eq:linearEoM},
for equilibrium magnetization along $z$ leads to the magnon dispersion,

 \begin{align}\label{magnon_disp}
 \omega_{\vec{q}}&=\frac{\mu_B}{M_{tot}}Im({J}^{xy}_{\vec{q}}-{J}^{yx}_{\vec{q}})\pm\omega_{\vec{q}}^0, 
 \end{align}	
 where, the first term represents the DMI contribution to the magnonic energy which is an odd function of $\vec{q}$, while the second term is an even function of $\vec{q}$ and is given by,
 \begin{align}
 \omega_{\vec{q}}^0&=\frac{\sqrt{4({J}^{xx}_{\vec{q}}-{J}^{zz}_{0})({J}^{yy}_{\vec{q}}-J^{zz}_{0})-({J}^{xy}_{\vec{q}}+{J}^{yx}_{\vec{q}})^2}}{M_{tot}/\mu_B}.
 \end{align}
 Similarly, using the equations of motion, Eq.~\eqref{eq:ContModel}, of the 
 micromagnetic model in the continuum limit, the magnon dispersion for a given equilibrium magnetization direction, $\vec{m}_{eq}$ is given by
 \begin{align}
 \omega_{\vec{q}}&=\frac{4V_M\mu_B}{M_{tot}}\sum_{\alpha\beta}D_{\alpha\beta}q^{\alpha}m_{eq}^{\beta}\pm\omega_{\vec{q}}^0, 
 \end{align}	
 where, the first term corresponds to the DMI contribution of the magnon energy and the second term is given by,
 \begin{align}
& \left(\frac{\omega_{\vec{q}}^0}{2\mu_BV_M}\right)^2 = (\vec{H}_{\vec{q}}^{eff}\cdot\vec{m}_{eq})^2+\sum_{\alpha\beta}adj[\hat{K}_{\vec{q}}^{eff}]_{\alpha\beta}m^{\alpha}_{eq}m^{\beta}_{eq}\nonumber\\
 &-\vec{H}_{\vec{q}}^{eff}\cdot\vec{m}_{eq}\sum_{\alpha}(K^{eff}_{\vec{q},\alpha\alpha}-\sum_{\beta}{K}^{eff}_{\vec{q},\alpha\beta}m^{\alpha}_{eq}m^{\beta}_{eq}).
 \end{align}	
 Here, $adj[...]$, represents the adjoint of the matrix and,
 \begin{align}
\vec{H}_{\vec{q}}^{eff}\cdot\vec{e}_{\alpha}&=\frac{2}{M_{tot}}\sum_{\beta}(A_{\alpha\beta}q_{\beta}^2m_{eq}^{\alpha}-K_{\alpha\beta}m_{eq}^{\beta})\\
{K}_{\vec{q},\alpha\beta}^{eff}&=\frac{2}{M_{tot}}\sum_{\gamma}(A_{\alpha\gamma}q_{\gamma}^2\delta_{\alpha\beta}-K_{\alpha\beta}).
 \end{align}	
 
 The equilibrium magnetization orientation, $\vec{m}_{eq}$ is found by the minimizing the magnetic energy of the collinear ferromagnet. 
 
Therefore, the DMI contribution to the magnon energy for any given equilibrium magnetization direction  in a system with isotropic crystal structure in x-y plane is given by,
\begin{align}\label{w_dmi}
\omega^{DMI}_{\vec{q}}=\frac{4V_M\mu_B}{M_{tot}}\Big(D^N\vec{e}_{z}\cdot\vec{q}\times\vec{m}_{eq}+D^{B}\vec{q}\cdot\vec{m}_{eq}\Big) .
\end{align}	
\subsection{First Principles Calculations of the Exchange Coupling}
We employ the Green's function method to calculate the exchange coupling elements. Within the linear combination 
of atomic orbitals (LCAO) basis, the multi-orbital Hamiltonian, $\hat{H}_{\vec{R}}$, describing the hopping of electrons between two unit-cells separated by $\vec{R}$, is given by,
\begin{align}\label{eq:Hamil}
\hat{H}_{\vec{R}}=\hat{H}^0_{\vec{R}}+\hat{H}^{soc}_{\vec{R}}+\hat{\Delta}_{\vec{R}}\hat{\sigma}_z.
\end{align}	
Here, $\hat{H}^0_{\vec{R}}$ is the paramagnetic (spin-independent) term, $\hat{H}^{soc}_{\vec{R}}$ is the SOC term and $\hat{\Delta}_{\vec{R}}$ is the exchange splitting matrix responsible for magnetism in the ferromagnet. Note that the SOC and exchange splitting terms are even and odd with respect to time reversal symmetry, respectively.

The effective magnetic field acting on the spin moment of the $i$th ion within a unit cell at $\vec{R}$ is given by
\begin{align}
\vec{B}_{\vec{R},i}&=-\Big\langle\frac{\partial \hat{\boldsymbol{H}}}{\partial \vec{m}^i_{\vec{R}}}\Big\rangle,
\end{align}	
where, $\langle...\rangle$ is the expectation value the Hamiltonian matrix elements, $\hat{\boldsymbol{H}}_{\vec{R},\vec{R}'}=\hat{H}_{\vec{R}-\vec{R}'}$.
The exchange coupling elements entering the magnon dispersion in Eq.~\eqref{magnon_disp} are calculated from,
\begin{subequations}
\label{eq:exchange}
\begin{align}
J_{0}^{i,zz}&=-\vec{B}^i_{0}\cdot\vec{e}_z\\
&=\int \frac{dE}{\pi} ImTr\Big(\hat{[\Delta G]}_{\vec{R}=0}\hat{\sigma}^z\hat{1}_i)\Big)f(E),\nonumber\\
J_{\vec{R}}^{ij,\alpha\beta}&=-\partial B^{i,\alpha}_{0}/\partial (\vec{m}^{j}_{\vec{R}}\cdot\vec{e}_{\beta})\\
&=\int \frac{dE}{\pi} Im Tr\Big(\hat{[\Delta G]}_{\vec{R}}\hat{1}_i\hat{\sigma}^{\alpha}\hat{[\Delta G]}_{-\vec{R}}\hat{\sigma}^{\beta}\hat{1}_{j}\Big)f(E)\nonumber.
\end{align}	
\end{subequations}
Here, $\hat{1}_i$ is the atomic position operator within the unit cell (diagonal matrix elements equal to one for atomic orbitals of the $i$th atom  and zero otherwise), $f(E)$ is the Fermi-Dirac distribution function and 
\begin{align}
\hat{G}_{\vec{k}}&=(E\hat{\mathcal{O}}_{\vec{k}}-\hat{H}_{\vec{k}})^{-1},
\end{align}	
is the Green's function, where the Hamiltonian, $\hat{\mathcal{H}}_{\vec{k}}$, 
and overlap, $\hat{\mathcal{O}}_{\vec{k}}$, matrices are calculated  
using the Linear Combination of Pseudo-Atomic Orbital (LCPAO) approach as implemented in the OpenMX package.\cite{OzakiPRB2003,OzakiPRB2004,OzakiPRB2005}

\section{Computational Details}\label{sec:DFT}
The Hamiltonian, $\hat{H}_{\vec{R}}$, and overlap, $\hat{\mathcal{O}}_{\vec{R}}$ matrix elements of the (111) Co/Pt slab consisting of various
thicknesses of Co and Pt  are determined from density functional theory calculations employing the 
OpenMX\cite{OzakiPRB2003,OzakiPRB2004,OzakiPRB2005} {\it ab initio} package.
We adopted Troullier-Martins type norm-conserving pseudopotentials\cite{TroullierPRB1991} with partial core correction. We used a $24\times 24\times 1$ k-point mesh for the first Brillouin zone (BZ) integration, and an energy cutoff of 500 Ry for numerical integrations in the real space grid. The localized orbitals were generated with radial cutoffs of 6.0 a.u and 7.0 a.u. for Co and Pt, respectively\cite{OzakiPRB2003,OzakiPRB2004}.
We used the in-plane lattice parameter of $a=2.7 \AA$   and the L(S)DA\cite{CeperleyPRL1980} exchange correlation functional as 
parameterized by Perdew and Zunger\cite{PerdewPRB1981}.
\begin{figure}
	{\includegraphics[scale=0.32,angle=0,trim={0cm 3.3cm 0.0cm 4.45cm},clip,width=0.5\textwidth]{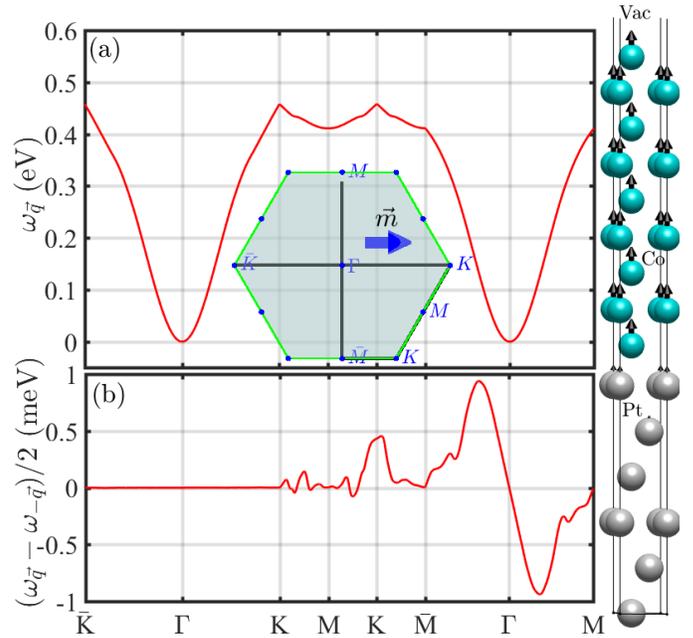}}%
	\caption{(a) Magnon dispersion in the spin-spiral regime along high symmetry directions in the first Brillouin zone, depicted in the inset, for the (111) Co(9 ML)/Pt(6 ML) bilayer with equilibrium magnetization along the $\bar{K}-K$ direction. (b) Anti-symmetric part of the magnon energy along high symmetry directions, where a non-zero slope around $\Gamma$ gives the DMI.}
	\label{fig:fig2}
\end{figure}

Structural relaxations were carried using the Vienna ab initio simulation package (VASP) \cite{Kresse96a,Kresse96b} within the generalized gradient approximation as parameterized by Perdew et al.\cite{PBE}
until the largest atomic force is smaller than 0.01 eV \AA$^{-1}$. The pseudopotential and wave functions are treated within the projector-augmented wave method \cite{Blochl94,KressePAW}.  A 15 \AA~ thick vacuum
region is introduced to separate the periodic slabs along the stacking direction. The plane wave cutoff energy was set to 500 eV and a 14 $\times$ 14 $\times$ 1 k-points mesh was used in the 2D BZ sampling. 

In the calculations of the exchange coupling the energy integration in Eq.~\eqref{eq:exchange} was carried out using the Matsubara summation approach with the poles obtained from Ozaki's continued fraction method of the Fermi-Dirac distribution function. \cite{Ozaki_CF_2007} The Fermi-Dirac distribution function temperature was set at $k_BT=25$ meV. 

\section{Results and Discussion}\label{sec:Results}

Fig.~\ref{fig:fig2}(a)  shows the magnon dispersion, $\omega_{\vec{q}}$, calculated from Eq.~\eqref{magnon_disp} along the high symmetry directions of the 2D BZ, shown in the inset, for the Pt(6 MLs)/Co(9 MLs) bilayer with in-plane magnetization along 
the $\bar{K}-K$ direction. The crystal structure is shown at the right side of Fig.~\ref{fig:fig2}. According to Eq.~\eqref{w_dmi}, the Bloch (N\'{e}el) DMI coefficient is given by the slope of $\omega^{DMI}_{\vec{q}}=(\omega_{\vec{q}}-\omega_{-\vec{q}})/2$ with respect to $q_y$ ($q_x$) around the $\Gamma$-point. 
Fig.~\ref{fig:fig2}(b) shows the DMI contribution to the magnon energy 
along the high symmetry directions. 
A finite DMI magnon energy along the $\bar{M}-M$ path $\vec{q}\perp\vec{m}_{eq}$ suggests the presence of a nonzero N\'{e}el  DMI, $D^N$, while the zero slope for the DMI magnon energy around $\Gamma$ along the $\vec{q}||\vec{m}$ direction demonstrates the absence of Bloch DMI, $D^B=0$.

\begin{figure} [tbp]
	{\includegraphics[scale=0.32,angle=0,trim={0cm 8.1cm 0.0cm 2.8cm},clip,width=0.5\textwidth]{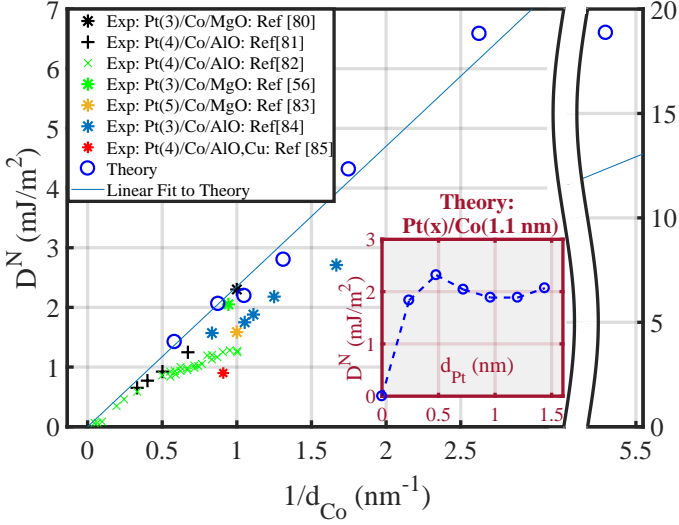}}%
	\caption {{\it Ab initio}  N\'{e}el  DMI (open blue circles) versus 
		inverse Co thicknesses for the (111) Pt (1.5 nm)/Co($d_{Co}$) bilayer slab along with their linear fit (blue line). 
		For comparison we also show the available experimental DMI values (stars and crosses) for various Pt($d_{Pt}$ nm)/Co(x) bilayers 
		grown on different insulating substrates.~\cite{Cao2018,Cho2018,Cho2015,Boulle2016,Ma2018,Belmeguenai2015,Kim2017}. 	
Inset: {\it Ab initio}  N\'{e}el  DMI versus Pt thicknesses for the Pt($d_{Pt}$)/Co (1.1 nm) bilayer.}
	\label{fig:fig3}
\end{figure}
Fig.~\ref{fig:fig3} displays the variation of the {\it ab initio} calculated N\'{e}el DMI (open blue circles) for the Pt(6 ML)/Co($d_{Co}$) bilayer as a function of the inverse Co thickness. The blue line is a linear square fit of the {\it ab initio} values. We also show for comparison the experimentally reported
DMI values (stars and crosses) ~\cite{Cao2018,Cho2018,Cho2015,Boulle2016,Ma2018,Belmeguenai2015,Kim2017} of Pt/Co(x)/insulator heterostructures with different insulators. 
Overall, we find a relatively good agreement between our results and previously
reported {\it ab initio} values\cite{Zimermann2019} as well as the experimentally reported data measured at room temperature. 
It should be noted that a small enhancement of the experimental results are expected due to the oxidation of the interfacial Co atoms with the insulating cap layer.\cite{Nembach2020}
The relatively small overestimation of the theoretical results, which is more significant for thinner Co films may be attributed to the effect of temperature
 and the diffusion of atoms near the interface\cite{Wells2017,Zimmerman2018,ChshievDMI2015}. 
The inset shows the dependence of the calculated N\'{e}el DMI on the Pt thicknesses for the Pt($d_{Pt})$/Co (1.1 nm) bilayer which is relatively weak, due to the fact that the interfacial Pt yields the large contributor to the DMI\cite{ChshievDMI2015}.  On the other hand, this result is in contrast to experimental observations \cite{Tacchi2017}, where a diffusion-like dependence of DMI versus Pt thickness has been reported. 

\begin{figure} [tbp]
	{\includegraphics[scale=0.32,angle=0,trim={0cm 9.4cm 0.0cm 3.0cm},clip,width=0.5\textwidth]{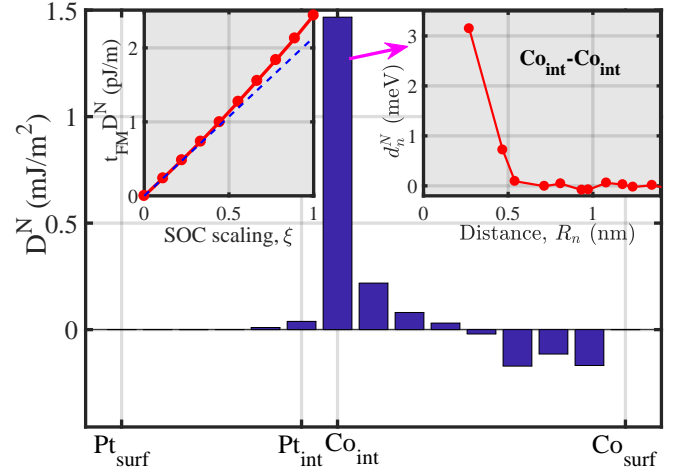}}%
	\caption{Layer-resolved N\'{e}el DMI for the Pt(6 MLs)/Co(9 MLs) bilayer 
		versus the layer index,where the Pt$_{int}$ and Co$_{int}$ denote the interfacial atoms while the Pt$_{surf}$ and Co$_{surf}$ denote the 
surface atoms.  Left Inset: Total interfacial DMI versus SOC scaling factor, where the dashed line denotes the linear variation of the DMI 
	with SOC. Right Inset: N\'{e}el DMI for a pair of Co atoms, $d^N_n$, 
on the interfacial layer versus their inter-atomic distance $R_n$.}
	\label{fig:fig4}
\end{figure}
In view of the interfacial nature of the  DMI in heavy metal/ferromagnet bilayers, it is expected that the interfacial Co atoms 
experience the dominant contribution of the DMI. From Eq.~\eqref{eq:ContEq} the DMI for an isotropic bilayer with magnetization direction along $y$ ($\vec{m}_{eq}||\vec{e}_y$)
and magnon propagation along $x$ can be decomposed into its layer- and atom-resolved contribution as following
\begin{align}\label{eq:Resolvd}
D^{N,ij}_{R_n}=\frac{1}{2V_M}\sum_{|\vec{R}|=R_n}J_{\vec{R}}^{ij,zx}\vec{R}\cdot\vec{e}_x.
\end{align}	
 In the case of the Pt/Co bilayer the atom-resolved DMI can be determined from 
$D^{N;i}=\sum_{j}(D^{N;ij}+D^{N;ji})/2$, where  $D^{N,ij}
  =\sum_{R_n}D^{N;ij}_{R_n}$. 

Fig.~\ref{fig:fig4} shows the layer-resolved DMI versus the layer index of 
the Pt(6 MLs)/Co(9 MLs) bilayer. As expected, we find that the interacial Co yields the dominant contribution. One can also determine from Eq.~\eqref{eq:Resolvd} the contribution of the $n$th nearest-neighbor atomic shell to the layer-resolved DMI. Comparing  Eq.~\eqref{eq:Resolvd}  with Eq.~\eqref{eq:PhenDMI} and the corresponding discussion, we obtain, 
\begin{align}
d^{N,ij}_n=\frac{4V_MD^{N;ij}_{R_n}}{N_nR_n}.
\end{align}	
The  right inset in Fig.~\ref{fig:fig4} shows the calculated, $d^{N,ij}_n$, where $i,j \in Co_{int}$ lie on the interfacial Co layer. We find that the dominant contribution arises from the first- and to a lesser extent the second-nearest neighbor atoms. The left inset in Fig.~\ref{fig:fig4} 
	displays the total interfacial DMI for Pt(6 MLs)/Co(9 MLs) versus the SOC scaling factor. The small enhancement of the DMI from its linear variation 
	(blue dashed line) for larger SOC values is due to the contributions from higher-order terms in SOC.
\begin{figure} [tbp]
	{\includegraphics[scale=0.32,angle=0,trim={0cm 13cm 0.0cm 1cm},clip,width=0.5\textwidth]{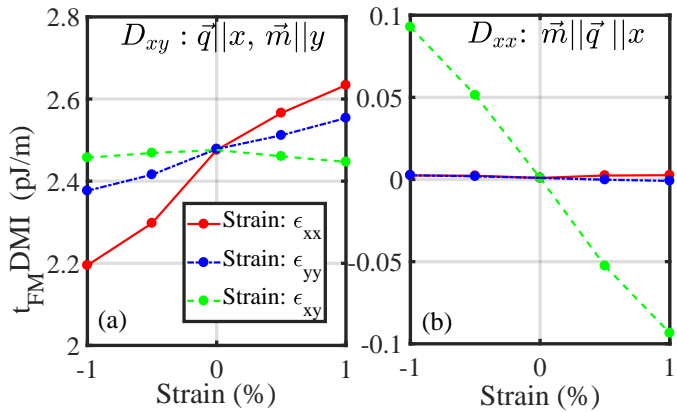}}%
	\caption{(a) Interfacial N\'{e}el DMI, $D_{xy}$,  of the 
Pt (6 ML)/Co (3 ML) bilayer for magnon propagation along $x$ and  magnetization direction along $y$ versus $\epsilon_{xx}$, $\epsilon_{xy}$ and  $\epsilon_{yy}$ strain;  
(b) Interfacial Bloch DMI, $D_{xx}$, for  magnon propagation and magnetization 
direction along $x$, versus $\epsilon_{xx}$, $\epsilon_{xy}$ and  $\epsilon_{yy}$ strain.}
	\label{fig:fig5}
\end{figure}

Finally, we have investigated the effect of strain 
on the DMI for various Pt and Co thickness. 
In the presence of strain, ${\epsilon_{ij}}$, the modified primitive lattice vectors, $\vec{a}'_i$, are given by $(\vec{a}'_i-\vec{a}_i)\cdot\vec{e}_j=\sum_{k}\epsilon_{kj}\vec{a}_{i}\cdot\vec{e}_k$, where the  $\vec{e}_j$'s represent 
the unit vectors in Cartesian coordinates. Note that in Voigt notation the off-diagonal elements of the strain matrix, $\epsilon_{i\neq j}$, should be divided by 2 to avoid double counting.\cite{Iotova1996,mahfouziME2020}
Fig.~\ref{fig:fig5}(a) and (b) show the N\'{e}el, $D_{yx}$, and Bloch, $D_{xx}$,
DMIs, respectively, multiplied by the thickness of the Co film, versus 
$\epsilon_{xx}$, $\epsilon_{yy}$ and $\epsilon_{xy}$, for the 
Pt (6 ML)/Co (3 ML) bilayer. 
We observe that under shear strain $\epsilon_{xy}$ non-vanishing 
{\it anisotropic} Bloch (diagonal) DMI components, $D_{xx}=-D_{yy}$, emerge, 
while the  N\'{e}el DMI, $(D_{xy},D_{yx})$ remains intact.

We have derived a general expression for the strain-induced change of the DMI given by,
\begin{align}
\delta D_{\alpha\beta}&=\Big[\frac{\epsilon_{\alpha\alpha}-\epsilon_{\beta\beta}}{2}\mathfrak{D}^N_{an}+\frac{\epsilon_{\alpha\alpha}+\epsilon_{\beta\beta}}{2}\mathfrak{D}_{iso}^{N}\Big]\vec{e}_{z}\cdot\vec{e}_{\alpha}\times\vec{e}_{\beta}\nonumber\\
&-2\mathfrak{D}^B_{an}\sum_{\alpha'\neq\alpha}\epsilon_{\alpha\alpha'}\vec{e}_{z}\cdot\vec{e}_{\alpha'}\times\vec{e}_{\beta},
\end{align}	
where the elasto-DMI coefficients are calculated from, 
$\mathfrak{D}^N_{iso}=\frac{\partial}{\partial\epsilon_{xx}}(D_{xy}-D_{yx})$, 
$\mathfrak{D}^N_{an}=\frac{\partial}{\partial\epsilon_{xx}}(D_{xy}+D_{yx})$ and 
$\mathfrak{D}^B_{an}=\frac{1}{2}\frac{\partial}{\partial\epsilon_{xy}}D_{xx}$. 
In Tabel.~\ref{table:tab1}, we list the values of the elasto-DMI coefficients 
for the Pt(6 MLs)/Co(3 MLs) and Pt(6 MLs)/Co(6 MLs) bilayers, respectively. 
The results show a larger modulation of the isotropic DMI, $\mathfrak{D}^N_{iso}$ under strain. Overall, the calculated elasto-DMI coefficients are in relative agreement with the recently reported experimental measurements.\cite{Gusev2020} 
Note that in polycrystalline systems an averaging of the anisotropic Bloch, $\mathfrak{D}^B_{an}$, and  N\'{e}el, $\mathfrak{D}^N_{an}$, components is expected. 
The opposite sign of the anisotropic DMI under shear strain, $\epsilon_{xy}$ compared to $\epsilon_{xx}$,  results in their cancellation in the polycrystalline Pt/Co bilayers.  
 
 \begin{table} [t]
 	\caption{ Calculated interfacial elasto-DMI coefficients,
 		$\mathfrak{D}^N_{iso}=\frac{\partial}{\partial\epsilon_{xx}}(D_{xy}-D_{yx})$, 
 		$\mathfrak{D}^N_{an}=\frac{\partial}{\partial\epsilon_{xx}}(D_{xy}+D_{yx})$ and 
 		$\mathfrak{D}^B_{an}=\frac{1}{2}\frac{\partial}{\partial\epsilon_{xy}}D_{xx}$		
 		in units of $(pJ/m)$ for the Pt(6 MLs)/Co(3 MLs) and Pt(6 MLs)/Co(6 MLs) bilayers, where $t_{FM}$ is the thickness of the Co thin film.}
 	\begin{ruledtabular}
 		\begin{tabular}{cccc}
 			&  $t_{FM}\mathfrak{D}^N_{iso} $  &  $t_{FM}\mathfrak{D}^N_{an} $  & $t_{FM}\mathfrak{D}^B_{an} $ 
 			\\
 			
 			\hline
 			Pt(6 MLs)/Co(3 MLs)
 			& 38  & 18 & -5  
 			\\
 			Pt(6 MLs)/Co(6 MLs)  &  52 & 14  &  -8
 			\\
 		\end{tabular}
 	\end{ruledtabular}\label{table:tab1}	
 	
 \end{table}
\section{Conclusions}\label{sec:Conclusions}
In summary, we have presented an accurate and computationally efficient approach to calculate the DMI from first principles calculations. The approach was applied to 
the (111) Pt/Co bilayer where the DMI was calculated for various thicknesses of the Co and Pt thin films. Overall, the {\it ab initio} results are in relatively good agreement with a wide range of experiments on different insulating substrates. 
The atom-resolved DMI suggests that the nearest-neighbor interfacial Co atoms yield the largest contribution to the DMI.  The  
 We have also investigated the modulation of the DMI under various types of strain which in turn gives rise to large anisotropic DMI components. We find large changes of the isotropic DMI component under $\epsilon_{\alpha\alpha}$ ($\alpha=x,y$) strain. 
 The strain-induced control of DMI in multiferroic devices may pave an efficient way for skyrmion and domain wall motion-based spintronic applications.

\section{ACKNOWLEDGMENTS}
	\begin{acknowledgments}
	The work is supported by NSF ERC-Translational Applications of Nanoscale Multiferroic Systems (TANMS)- Grant No. 1160504, NSF PFI-RP Grant No. 1919109,
	and by NSF-Partnership in Research and Education in Materials (PREM) Grant No. 
	DMR-1828019.
	\end{acknowledgments}

\appendix


	

\begin{thebibliography}{10}
		
		
		\bibitem{Fert2013}	
		 A. Fert, V. Cros, J. Sampaio, Skyrmions on the track. Nature Nanotechnology {\bf 8}(3), 152–156 (2013).
		\bibitem{Kiselev2011}	
		N. Kiselev, A. Bogdanov, R. Schfer, U. Rler, Chiral skyrmions in thin magnetic films: new objects for magnetic storage
		technologies? Journal of Physics D: Applied Physics {\bf 44}, 392001 (2011).
		\bibitem{Nagaosa2013}	
		N. Nagaosa, Y. Tokura, Topological properties and dynamics of magnetic skyrmions. Nature Nanotechnology {\bf 8}(12), 899–911
		(2013).
		
		
		\bibitem{Bogdanov1989}	
		A. N.  Bogdanov, D. A.  Yablonskii, Thermodynamically stable vortices in magnetically ordered crystals. Te mixed state of magnets.
		Zh. Eksp. Teor. Fiz {\bf 95}(1), 178 (1989).
		\bibitem{Bogdanov1994}	
		 A.  Bogdanov,  A. Hubert, Thermodynamically stable magnetic vortex states in magnetic crystals. Journal of Magnetism and
		Magnetic Materials {\bf 138}(3), 255–269 (1994).
		
		\bibitem{Muhlbauer2009}
		 S. M{\"u}hlbauer, B.  Binz, F.  Jonietz, C.  Pfleiderer, A. Rosch, A.  Neubauer, R.  Georgii, and P. B{\"o}ni, 
		 Skyrmion Lattice in a Chiral Magnet,
		 Science {\bf 323}, 915 (2009).
		\bibitem{Yu2010}
		X. Z. Yu, Y. Onose, N. Kanazawa, J. H. Park, J. H. Han, Y. Matsui, N. Nagaosa and Y. Tokura , Real-space observation of a two-dimensional skyrmion crystal. Nature {\bf 465}, 901–904 (2010).
		\bibitem{Yu2011}
		X. Z. Yu, N. Kanazawa, Y. Onose, K. Kimoto, W. Z. Zhang, S. Ishiwata, Y. Matsui and Y. Tokura , Near room-temperature formation of a skyrmion crystal in thin-flms of the helimagnet FeGe. Nature Materials {\bf 10}(2),
		106–109 (2011).
		\bibitem{Shibata2013}
		K. Shibata, X. Z. Yu, T. Hara, D. Morikawa, N. Kanazawa, K. Kimoto, S. Ishiwata, Y. Matsui and Y. Tokura, Towards control of the size and helicity of skyrmions in helimagnetic alloys by spin - orbit coupling. Nature
		Nanotechnology {\bf 8}, 723–728 (2013).
		
		
		
		\bibitem{Fert2017}
		A. Fert, N. Reyren, and V. Cros, 
		Magnetic skyrmions: advances in physics and potential applications,
		Nat. Rev. Mater. {\bf 2}, 17031, (2017).
		\bibitem{Everschor2018}
		K. Everschor-Sitte, J. Masell, R. M. Reeve, and M. Klui, 
		Perspective: Magnetic skyrmions—Overview of recent progress in an active research field,
		J. Appl. Phys. {\bf 124}, 240901 (2018).		
		\bibitem{Finocchio2016}
		G. Finocchio, F. Bttner, R. Tomasello, M. Carpentieri, and	M. Klui, 
		Magnetic skyrmions: from fundamental to applications, 
		J. Phys. D {\bf 49}, 423001 (2016).
		
		\bibitem{Jiang2015}
		W. Jiang, P. Upadhyaya, W. Zhang, G. Yu, M. B.	Jungfleisch, F. Y. Fradin, J. E. Pearson, Y. Tserkovnyak,	K. L. Wang, O. Heinonen, S. G. E. te Velthuis, and A. Hoffmann, 
		Blowing magnetic skyrmion bubbles,
		Science {\bf 349}, 283 (2015).
		\bibitem{Hrabec2017}
		A. Hrabec, J. Sampaio, M. Belmeguenai, I. Gross, R. Weil,	S. M. Chrif, A. Stashkevich, V. Jacques, A. Thiaville, and	S. Rohart, 
		Current-induced skyrmion generation and dynamics in symmetric bilayers,
		Nat. Commun. {\bf 8}, 15765 (2017).
		\bibitem{Sampaio2013}
		J. Sampaio, V. Cros, S. Rohart, A. Thiaville, and A. Fert,	
		Nucleation, stability and current-induced motion of isolated magnetic skyrmions in nanostructures,
		Nat. Nanotechnol. 8, {\bf 839} (2013).
		\bibitem{Upadhyaya2016}
		G. Yu, P. Upadhyaya, X. Li, W. Li, S. K. Kim, Y. Fan, K. L.	Wong, Y. Tserkovnyak, P. K. Amiri, and K. L. Wang, 
		Room-Temperature Creation and Spin–Orbit Torque Manipulation of Skyrmions in Thin Films with Engineered Asymmetry,
		Nano	Lett. {\bf 16}, 1981 (2016).
		\bibitem{Bttner2017}
		F. Bttner, I. Lemesh, M. Schneider, B. Pfau, C. M. Gnther,	P. Hessing, J. Geilhufe, L. Caretta, D. Engel, B. Krger, J. Viefhaus, S. Eisebitt, and G. S. D. Beach, 
		Field-free deterministic ultrafast creation of magnetic skyrmions by spin–orbit torques,
		Nat. Nanotechnol. {\bf 12}, 1040 (2017).	
		
		\bibitem{HellmanRMP2017}
		F. Hellman, A. Hoffmann, Y. Tserkovnyak, G. S. D. Beach,  E. E. Fullerton,  C. Leighton, A. H. MacDonald, D. C. Ralph, D. A. Arena, H. A. Durr, P. Fischer, J. Grollier, J. P. Heremans, T. Jungwirth, A. V. Kimel, B. Koopmans, I. N. Krivorotov, S.J. May, A.K. Petford-Long, J.M. Rondinelli, N. Samarth, I.K. Schuller, A.N. Slavin,M.D. Stiles, O. Tchernyshyov, A. Thiaville, B.L. Zink,
		Interface-induced phenomena in magnetism,
		Rev. Mod. Phys. {\bf 89}, 025006 (2017).	
		\bibitem{Soumyanarayanan2017}
		A. Soumyanarayanan, M. Raju, A. L. Gonzalez Oyarce,	A. K. C. Tan, M.-Y. Im, A. P. Petrovic, P. Ho, K. H. Khoo,	M. Tran, C. K. Gan, F. Ernult, and C. Panagopoulos, 
		Tunable room-temperature magnetic skyrmions in Ir/Fe/Co/Pt multilayers. 
		Nat. Mater. {\bf 16}, 898 (2017).
				
		\bibitem{Del1958}	
		 I. J. Dzyaloshinsky, 
		 A thermodynamic theory of “weak” ferromagnetism of antiferromagnetics,
		 Phys. Chem. Solids {\bf 4}, 241 (1958).
		\bibitem{Moriya1960}	
		 T. Moriya, 
		 Anisotropic Superexchange Interaction and Weak Ferromagnetism,
		 Phys. Rev. {\bf 120}, 91 (1960) 
		 
		 
		 		 
		 
		 \bibitem{Udvardi2009}	
		 L. Udvardi and L. Szunyogh, 
		 Chiral Asymmetry of the Spin-Wave Spectra in Ultrathin Magnetic Films,
		 Phys. Rev. Lett. {\bf 102}, 207204 (2009).
		\bibitem{Costa2010}	
		A. T. Costa, R. B. Muniz, S. Lounis, A. B. Klautau, and	D. L. Mills, 
		Spin-orbit coupling and spin waves in ultrathin ferromagnets: The spin-wave Rashba effect,
		Phys. Rev. B {\bf 82}, 014428 (2010).
		\bibitem{Moon2013}	
		J. H. Moon, S. M. Seo, K. J. Lee, K. W. Kim, J. Ryu, H. W. Lee, R. D. McMichael, and M. D. Stiles, 
		Spin-wave propagation in the presence of interfacial Dzyaloshinskii-Moriya interaction,
		Phys. Rev. B 88, 184404 (2013).
		\bibitem{Volkov2019}	
		 A. F. Volkov, F. S. Bergeret, and K. B. Efetov, 
		 Spin polarization and orbital effects in superconductor-ferromagnet structures,
		 Phys. Rev. B {\bf 99}, 144506 (2019).
		\bibitem{Di2015}	
		K. Di, V. L. Zhang, H. S. Lim, S. C. Ng, M. H. Kuok, J. Yu, J. Yoon, X. Qiu, and H. Yang, 
		Direct Observation of the Dzyaloshinskii-Moriya Interaction in a Pt/Co/Ni Film,
		Phys. Rev. Lett. {\bf 114}, 047201 (2015).
		\bibitem{Ma2016}	
		 X. Ma, G. Yu, X. Li, T. Wang, D. Wu, K. S. Olsson, Z. Chu, K. An, J. Q. Xiao, K. L. Wang, and X. Li,
		 Interfacial control of Dzyaloshinskii-Moriya interaction in heavy metal/ferromagnetic metal thin film heterostructures,
		Phys. Rev. B {\bf 94}, 180408(R) (2016).
		\bibitem{Ma2017}	
		 X. Ma, G. Yu, S. A. Razavi, S. S. Sasaki, X. Li, K. Hao, S. H. Tolbert, K. L. Wang, and X. Li, 
		 Dzyaloshinskii-Moriya Interaction across an Antiferromagnet-Ferromagnet Interface,
		Phys. Rev. Lett. {\bf 119}, 027202 (2017).
		
		\bibitem{Kim2018}
		Kyoung-Whan Kim, Kyoung-Woong Moon,2 Nico Kerber, Jonas Nothhelfer, and Karin Everschor-Sitte, 
		Asymmetric skyrmion Hall effect in systems with a hybrid Dzyaloshinskii-Moriya interaction,
		Phys. Rev. B {\bf 97}, 224427 (2018).
		\bibitem{Yang2018}
		Hongxin Yang, Gong Chen, Alexandre A. C. Cotta, Alpha T. NDiaye, Sergey A. Nikolaev, Edmar A. Soares, Waldemar A. A. Macedo, Kai Liu, Andreas K. Schmid, Albert Fert and Mairbek Chshiev 
		Significant Dzyaloshinskii-Moriya interaction at graphene–ferromagnet interfaces due to the Rashba effect,
		 Nat. Mater. {\bf 17}, 605 (2018).
		\bibitem{Manchon2015}
		A. Manchon, H. C. Koo, J. Nitta, S. Frolov and R. Duine, 
		New perspectives for rashba spin–orbit coupling,
		Nat. Mater. {\bf 14}, 871 (2015).		
		\bibitem{Kezsmarki2015}
		I. Kézsmárki, S. Bordács, P. Milde, E. Neuber, L. M. Eng, J. S. White, H. M. Rønnow, C. D. Dewhurst, M. Mochizuki, K. Yanai, H. Nakamura, D. Ehlers, V. Tsurkan and A. Loidl, 
		Néel-type skyrmion lattice with confined orientation in the polar magnetic semiconductor GaV4S8
		Nat. Mater. {\bf 14}, 1116 (2015).
		\bibitem{Kurumaji2017}
		T. Kurumaji, T. Nakajima, V. Ukleev, A. Feoktystov, T.-h. Arima, K. Kakurai, and Y. Tokura, 
		Néel-Type Skyrmion Lattice in the Tetragonal Polar Magnet 
		VOSe2O5,
		Phys. Rev. Lett. {\bf 119}, 237201 (2017).
		
		
		\bibitem{Onose2010}
		X. Yu, Y. Onose, N. Kanazawa, J. Park, J. Han, Y. Matsui, N. Nagaosa, and Y. Tokura, 
		Real-space observation of a two-dimensional skyrmion crystal,
		Nature (London) {\bf 465}, 901 (2010).
		\bibitem{Seki2010}
		S. Seki, X. Yu, S. Ishiwata, and Y. Tokura, 
		Observation of Skyrmions in a Multiferroic Material,
		Science {\bf 336}, 198 (2012).
		
		\bibitem{Finocchio2019}
		G. Finocchio, M. Di Ventra, K.Y. Camsari, K. Everschor-Sitte, P. K. Amiri, Z. Zeng,
		The promise of spintronics for unconventional computing. 
		Journal of Magnetism and Magnetic Materials,
		{\bf 521}, 167506 (2021).
		
		
		
		
		\bibitem{Tomasello2014}
		R. Tomasello, E. Martinez, R. Zivieri, L. Torres, M. Carpentieri and G. Finocchio,
		A strategy for the design of skyrmion racetrack memories,
		Scientific Reports {\bf 4}, 6784 (2014).
		\bibitem{Kang2016}
		Wang Kang, Yangqi Huang, Chentian Zheng, Weifeng Lv, Na Lei, Youguang Zhang, Xichao Zhang, Yan Zhou and Weisheng Zhao,  
		Voltage Controlled Magnetic Skyrmion Motion for Racetrack Memory,
		Scientific Reports {\bf 6}, 23164 (2016).
		\bibitem{Zhang2015_1}
		Xichao Zhang, Yan Zhou, Motohiko Ezawa, G. P. Zhao and Weisheng Zhao, 
		Magnetic skyrmion transistor: skyrmion motion in a voltage-gated nanotrack. 
		Scientific Reports {\bf 5}, 11369, (2015).
		\bibitem{Zhang2015_2}
		X. Zhang, M.  Ezawa, Y. Zhou, 
		Magnetic skyrmion logic gates: conversion, duplication and merging of skyrmions,
		Sci. Rep. {\bf 5}, 9400 (2015).
		
		\bibitem{Finocchio2015}
		G. Finocchio, M. Ricci, R. Tomasello, A. Giordano, M. Lanuzza, V. Puliafito, P. Burrascano,  B. Azzerboni, and M. Carpentieri, 
		Skyrmion based microwave detectors and harvesting,
		Appl. Phys. Lett. {\bf 107}, 262401 (2015). 
		
		\bibitem{He2017}
		Z. He, D. Fan, 
		A Tunable Magnetic Skyrmion Neuron Cluster for Energy Efficient Artificial Neural Network. In 2017 Design,
		Automation, Test in Europe Conference, Exhibition, 350–355 (2017).
		
		
		\bibitem{Li2018}
		Sai Li; Wang Kang; Xing Chen; Jinyu Bai; Biao Pan; Youguang Zhang; Weisheng Zhao,
		Emerging Neuromorphic Computing Paradigms Exploring Magnetic Skyrmions. in 2018 IEEE Computer Society Annual Symposium on VLSI (ISVLSI) 539–544 (IEEE, 2018). 
		
		\bibitem{Romming2015}
		Niklas Romming, André Kubetzka, Christian Hanneken, Kirsten von Bergmann, and Roland Wiesendanger,
		Field-Dependent Size and Shape of Single Magnetic Skyrmions,
		Phys. Rev. Lett. {\bf 114}, 177203 (2015).
		\bibitem{Koshibae2014}
		W. Koshibae, N. Nagaosa, 
		Creation of skyrmions and antiskyrmions by local heating, 
		Nature Communications {\bf 5}, 5148, (2014).
		
		\bibitem{Rowland2015}
		J. Rowland, S. Banerjee, M.  Randeria, 
		Skyrmions in Chiral Magnets with Rashba and Dresselhaus Spin-Orbit Coupling, 
		Phys. Rev. B {\bf 93}, 020404(R) (2016).
		
		\bibitem{Daz2016}
		S. A. Daz, R. E. Troncoso, 
		Controlling skyrmion helicity via engineered Dzyaloshinskii-Moriya interactions,
		J. Phys.: Condens. Matter {\bf 28}, 426005 (2016).
		
		\bibitem{Masell2020}
		Jan Masell, Karin Everschor-Sitte, 
		Current-Induced Dynamics of Chiral Magnetic Structures: Creation, Motion, and Applications, 
		ArXiv pre-print, 2004.13535 (2020).
		
		\bibitem{Buttner2017}
		Felix Büttner, Ivan Lemesh, Michael Schneider, Bastian Pfau, Christian M. Günther, Piet Hessing, Jan Geilhufe, Lucas Caretta, Dieter Engel, Benjamin Krüger, Jens Viefhaus, Stefan Eisebitt and Geoffrey S. D. Beach, 
		Field-free deterministic ultrafast creation of magnetic skyrmions by spin-orbit torques,
		Nat. Nanotechnol. {\bf 12}, 1040–1044 (2017). 
		
		
		\bibitem{Koyama2018}
		Tomohiro Koyama, Yoshinobu Nakatani, Jun’ichi Ieda, and Daichi Chiba, 
		Electric field control of magnetic domain wall motion via modulation of the Dzyaloshinskii-Moriya interaction, 
		Sci. Adv. {\bf 4}: eaav0265 (2018).
		\bibitem{Zhang2018}
		W. Zhang, H. Zhong, R. Zang, Y. Zhang, S. Yu, G. Han, G. L. Liu, S. S. Yan , S. Kang, and L. M. Mei,  
		Electrical field enhanced interfacial Dzyaloshinskii-Moriya interaction in MgO/Fe/Pt system, 
		Appl. Phys. Lett. {\bf 113}, 122406 (2018).
		\bibitem{Srivastava2018}
		T. Srivastava, M. Schott, R. Juge, V. Krizzakova, M. Belmeguenai,
		Y. Roussigne, A. Bernand-Mantel, L. Ranno, S. Pizzini, S.-M. Cherif, A. Stashkevich, S. Auffret, O. Boulle, G. Gaudin, M. Chshiev, C. Baraduc, and H. Bea, 
		Large-Voltage Tuning of Dzyaloshinskii-Moriya Interaction: A Route toward Dynamic Control of Skyrmion Chirality,  
		Nano Lett. {\bf 18}, 4871-4877 (2018).
		
		
		\bibitem{Gusev2020}
		N. S. Gusev, A. V. Sadovnikov, S. A. Nikitov, M. V. Sapozhnikov, and O. G. Udalov, Manipulation of the Dzyaloshinskii–Moriya Interaction in Co/Pt Multilayers with Strain, 
		Phys. Rev. Lett. {\bf 124}, 157202 (2020).
		
		
		
		
		
		
		\bibitem{Woo2016}
		Seonghoon Woo, Kai Litzius, Benjamin Krüger, Mi-Young Im, Lucas Caretta, Kornel Richter, Maxwell Mann, Andrea Krone, Robert M. Reeve, Markus Weigand, Parnika Agrawal, Ivan Lemesh, Mohamad-Assaad Mawass, Peter Fischer, Mathias Kläui and Geoffrey S. D. Beach,
		Observation of room-temperature magnetic skyrmions and their current-driven dynamics in ultrathin metallic ferromagnets,
		Nature Materials, {\bf 15}, 501–506 (2016).	
		
		\bibitem{Moreau2016}
		C. Moreau-Luchaire, C. Moutafis, N. Reyren, J. Sampaio, C. A. F. Vaz, N. Van Horne, K. Bouzehouane, K. Garcia, C. Deranlot, P. Warnicke, P. Wohlhüter, J.-M. George, M. Weigand, J. Raabe, V. Cros and A. Fert, 
		Additive interfacial chiral interaction in multilayers for stabilization of small individual skyrmions at room temperature,
		Nature Nanotechnology {\bf 11}, pages444–448(2016)
	
		\bibitem{Boulle2016}
		Olivier Boulle, Jan Vogel, Hongxin Yang, Stefania Pizzini, Dayane de Souza Chaves, Andrea Locatelli, Tevfik Onur Menteş, Alessandro Sala, Liliana D. Buda-Prejbeanu, Olivier Klein, Mohamed Belmeguenai, Yves Roussigné, Andrey Stashkevich, Salim Mourad Chérif, Lucia Aballe, Michael Foerster, Mairbek Chshiev, Stéphane Auffret, Ioan Mihai Miron and Gilles Gaudin, 
		Room-temperature chiral magnetic skyrmions in ultrathin magnetic nanostructures,
		Nature Nanotechnology, {\bf 11}, 449–454 (2016).
		
		
		
		\bibitem{ChshievDMI2015}
		Hongxin Yang, Andre Thiaville, Stanislas Rohart, Albert Fert, and Mairbek Chshiev, 
		Anatomy of Dzyaloshinskii-Moriya Interaction at Co/Pt Interfaces,
		Phys. Rev. Lett. {\bf 115}, 267210 (2015).
		
		
		\bibitem{Zimermann2019}
		Bernd Zimmermann, Gustav Bihlmayer, Marie Böttcher, Mohammed Bouhassoune, Samir Lounis,
		Jairo Sinova, Stefan Heinze, Stefan Blügel, and Bertrand Dupe, 
		Comparison of first-principles methods to extract magnetic parameters in ultrathin films: Co/Pt(111), 
		Phys. Rev. B {\bf 99}, 214426 (2019).
		
		\bibitem{Mankowsky2017}
		S. Mankovsky and H. Ebert,
		Accurate scheme to calculate the interatomic Dzyaloshinskii-Moriya interaction parameters,
		Phys. Rev. B {\bf 96}, 104416 (2017).
		
		
		
		\bibitem{Frank2017}
		Frank Freimuth, Stefan Blügel, and Yuriy Mokrousov,
		Relation of the Dzyaloshinskii-Moriya interaction to spin currents and to the spin-orbit field,
		Phys. Rev. B {\bf 96}, 054403 (2017).
		
		

\bibitem{Kato2019}
Naoaki Kato, Masashi Kawaguchi, Yong-Chang Lau, Toru Kikuchi, Yoshinobu Nakatani, and Masamitsu Hayashi,
Current-Induced Modulation of the Interfacial Dzyaloshinskii-Moriya Interaction,
Phys. Rev. Lett. {\bf 122}, 257205 (2019).

\bibitem{Yang2018_1}
Hongxin Yang, Olivier Boulle, Vincent Cros, Albert Fert and Mairbek Chshiev,
Controlling Dzyaloshinskii-Moriya Interaction via Chirality Dependent Atomic-Layer Stacking, Insulator Capping and Electric Field,
Scientific Reports {\bf 8}, 12356 (2018).

\bibitem{Deger2020}
Caner Deger,
Strain-enhanced Dzyaloshinskii–Moriya interaction at Co/Pt interfaces,
Scientific Reports {\bf 10}, 12314 (2020).	

\bibitem{Udalov2020}
Strain-dependent Dzyaloshinskii-Moriya interaction in a ferromagnet/heavy-metal bilayer,
O. G. Udalov 1,2,3 and I. S. Beloborodov,
Phys. Rev. B {\bf 102}, 134422 (2020).
		
\bibitem{Ado2020}	
I. A. Ado, A. Qaiumzadeh, A. Brataas, and M. Titov,
Chiral ferromagnetism beyond Lifshitz invariants,
Phys. Rev. B {\bf 101}, 161403(R) (2020).
	
	\bibitem{Camosi}
	Lorenzo Camosi, Stanislas Rohart, Olivier Fruchart, Stefania Pizzini, Mohamed Belmeguenai, Yves Roussigné, Andreï Stashkevich, Salim Mourad Cherif, Laurent Ranno, Maurizio de Santis, and Jan Vogel,
	Anisotropic Dzyaloshinskii-Moriya Interaction in ultra-thin epitaxial Au/Co/W(110),
	Phys. Rev. B {\bf 95}, 214422 (2017)
		
	
	\bibitem{Zimmermann2017}
	Markus Hoffmann, Bernd Zimmermann, Gideon P. Müller, Daniel Schürhoff, Nikolai S. Kiselev, Christof Melcher and Stefan Blugel,
	Antiskyrmions stabilized at interfaces by anisotropic Dzyaloshinskii-Moriya interactions,
	Nature Communications {\bf 8}, 308 (2017).	
		
		
		\bibitem{OzakiPRB2003}
		T. Ozaki, 
		Variationally optimized atomic orbitals for large-scale electronic structures,
		Phys. Rev. B {\bf 67}, 155108 (2003).		
		\bibitem{OzakiPRB2004}
		T. Ozaki, H. Kino, 
		Numerical atomic basis orbitals from H to Kr,
		Phys. Rev. B {\bf 69}, 195113 (2004).
		\bibitem{OzakiPRB2005}
		T. Ozaki, H. Kino, 
		Efficient projector expansion for the ab initio LCAO method,
		Phys. Rev. B {\bf 72}, 045121 (2005).
		\bibitem{TroullierPRB1991}
		N. Troullier, J. L. Martins, 
		Efficient pseudopotentials for plane-wave calculations,
		Phys. Rev. B {\bf 43}, 1993 (1991).
		
		\bibitem{CeperleyPRL1980}
		D. M. Ceperley, B. J. Alder, 
		Ground state of the electron gas by a stochastic method,
		Phys. Rev. Lett. {\bf 45}, 566 (1980).
		\bibitem{PerdewPRB1981}
		J. P. Perdew, A.  Zunger,
		Self-interaction correction to density-functional approximations for many-electron systems,
		Phys. Rev. B {\bf 23}, 5048 (1981). 	
		
		
		\bibitem{Kresse96a} 
		G. Kresse and J. Furthm\"uller, 
		Efficient iterative schemes for ab initio total-energy calculations using a plane-wave basis set,
		Phys. Rev. B \textbf{54}, 11169 (1996).
		\bibitem{Kresse96b}  
		G. Kresse and J. Furthm\"uller, 
		Efficiency of ab-initio total energy calculations for metals and semiconductors using a plane-wave basis set,
		Comput. Mater. Sci. \textbf{6}, 15 (1996).
		\bibitem{PBE} 
		J. P. Perdew, K. Burke, and M. Ernzerhof, 
		Generalized Gradient Approximation Made Simple,
		Phys. Rev. Lett. \textbf{77}, 3865 (1996).	
		\bibitem{KressePAW} 
		G. Kresse and D. Joubert, 
		From ultrasoft pseudopotentials to the projector augmented-wave method,
		Phys. Rev. B \textbf{59}, 1758 (1999).
		\bibitem{Blochl94} 
		P. E. Bl\"ochl, 
		Projector augmented-wave method,
		Phys. Rev. B \textbf{50}, 17953 (1994).	
		
		
		\bibitem{Ozaki_CF_2007}
		Taisuke Ozaki, 
		Continued fraction representation of the Fermi-Dirac function for large-scale electronic structure calculations,
		Phys. Rev. B {\bf 75}, 035123 (2007).
		
		
		\bibitem{Cao2018}
		Anni Cao, Xueying Zhang, Bert Koopmans,   Shouzhong Peng,   Yu Zhang,   Zilu Wang,   Shaohua Yan,   Hongxin Yang and  Weisheng Zhao,
		Tuning the Dzyaloshinskii–Moriya interaction in Pt/Co/MgO heterostructures through the MgO thickness, 
		Nanoscale, {\bf 10}, 12062-12067, (2018).
		
		\bibitem{Cho2018}
		Jaehun Cho, Nam-Hui Kim, Jinyong Jung, Dong-Soo Han, Henk J. M. Swagten, June-Seo Kim, and Chun-Yeol You, 
		Thermal Annealing Effects on the Interfacial Dzyaloshinskii–Moriya Interaction Energy Density and Perpendicular Magnetic Anisotropy,
		IEEE TRANSACTIONS ON MAGNETICS, {\bf 54}, 6 (2018).
		
		\bibitem{Cho2015}
		Jaehun Cho, Nam-Hui Kim, Sukmock Lee, June-Seo Kim, Reinoud Lavrijsen, Aurelie Solignac, Yuxiang Yin, Dong-Soo Han, Niels J. J. van Hoof, Henk J. M. Swagten, Bert Koopmans and  Chun-Yeol You,
		Thickness dependence of the interfacial Dzyaloshinskii–Moriya interaction in inversion symmetry broken systems, 
		Nature Communications volume {\bf 6}, 7635 (2015).
		
		
		\bibitem{Ma2018}
		Xin Ma, Guoqiang Yu, Chi Tang, Xiang Li, Congli He, Jing Shi, Kang L. Wang, and Xiaoqin Li,		
		Interfacial Dzyaloshinskii-Moriya Interaction: Effect of 5d	Band Filling and Correlation with Spin Mixing Conductance,
		Phys. Rev. Lett. {\bf 120}, 157204 (2018).
		
		\bibitem{Belmeguenai2015}
		Mohamed Belmeguenai, Jean-Paul Adam, Yves Roussigné, Sylvain Eimer, Thibaut Devolder, Joo-Von Kim, Salim Mourad Cherif, Andrey Stashkevich, and André Thiaville,
		Interfacial Dzyaloshinskii-Moriya interaction in perpendicularly magnetized Pt/Co/AlO$_x$ ultrathin films measured by Brillouin light spectroscopy,
		Phys. Rev. B {\bf 91}, 180405(R) (2015).
		
    	\bibitem{Kim2017}
		Nam-Hui Kim,  Jaehun Cho, Jinyong Jung, Dong-Soo Han,  Yuxiang Yin, June-Seo Kim, Henk J. M. Swagten, Kyujoon Lee, Myung-Hwa Jung, and  Chun-Yeol You,
		Role of top and bottom interfaces of a Pt/Co/AlOx system in Dzyaloshinskii-Moriya interaction, interface perpendicular magnetic anisotropy, and magneto-optical Kerr effect
		AIP Advances {\bf 7}, 035213 (2017)
		
		
		
		
		\bibitem{Nembach2020}
		Hans T. Nembach, Emilie Jue, Eric R. Evarts, and Justin M. Shaw,
		Correlation between Dzyaloshinskii-Moriya interaction and orbital angular momentum at an oxide-ferromagnet interface,
		Phys. Rev. B {\bf 101}, 020409(R) (2020).
		
		
		
		
		
		
		
		
		
		\bibitem{Wells2017}
		Adam W. J. Wells, Philippa M. Shepley, Christopher H. Marrows, and Thomas A. Moore, 
		Effect of interfacial intermixing on the Dzyaloshinskii-Moriya interaction in Pt/Co/Pt,
		Phys. Rev. B {\bf 95}, 054428 (2017).
		
		\bibitem{Zimmerman2018}
		Bernd Zimmermann , William Legrand , Davide Maccariello , Nicolas Reyren , Vincent Cros, Stefan Blügel, and Albert Fert, 
		Dzyaloshinskii-Moriya interaction at disordered interfaces from ab initio theory: Robustness against intermixing and tunability through dusting, 
		Appl. Phys. Lett. {\bf 113}, 232403 (2018). 
		
		
		
		
		
		\bibitem{Tacchi2017}
		S. Tacchi, R. E. Troncoso, M. Ahlberg, G. Gubbiotti, M. Madami, J. Akerman, and P. Landeros, 
		Interfacial Dzyaloshinskii-Moriya Interaction in Pt/CoFeB Films: Effect of the Heavy-Metal Thickness,
		Phys. Rev. Lett. {\bf 118}, 147201 (2017).
		\bibitem{Iotova1996}
		Dobrina Iotova, Nicholas Kioussis, and Say Peng Lim,		
		Electronic structure and elastic properties of the Ni3X (X=Mn, Al, Ga, Si, Ge) intermetallics,
		Phys. Rev. B {\bf 54}, 14413 (1996).
		
		\bibitem{mahfouziME2020}
		Farzad Mahfouzi, Gregory P. Carman, and Nicholas Kioussis,
		Magnetoelastic and magnetostrictive properties of Co2XAl Heusler compounds,
		Phys. Rev. B {\bf 102}, 094401 (2020).
		

		
					
	
		
		
		
		
		
		
		
		
	\end{thebibliography}
	
\end{document}